\documentclass[]{aastex631}
\usepackage{amsmath} 

\begin{document}

\title{Simulating the Arrival of Multiple Coronal Mass Ejections that Triggered the Gannon Superstorm on May 10, 2024}

\author{Smitha V. Thampi}
\affiliation{Space Physics Laboratory, Vikram Sarabhai Space Centre, Thiruvananthapuram, 695022, 
India}
\author{Ankush Bhaskar}
\affiliation{Space Physics Laboratory, Vikram Sarabhai Space Centre, Thiruvananthapuram, 695022, 
India}

\author{Prateek Mayank}
\affiliation{Department of Astronomy, Astrophysics and Space Engineering, Indian Institute of Technology-Indore, Madhya Pradesh, 453552, India.}

\author{Bhargav Vaidya}
\affiliation{Department of Astronomy, Astrophysics and Space Engineering, Indian Institute of Technology-Indore, Madhya Pradesh, 453552, India.}

\author{Indu Venugopal}

\affiliation{Space Physics Laboratory, Vikram Sarabhai Space Centre, Thiruvananthapuram, 695022, 
India}

\affiliation{Department of Physics, CUSAT, Kochi, 682022, India}


\begin{abstract}

The May 10, 2024 space weather event stands out as the most powerful storm recorded during the current solar cycle. This study employs a numerical framework utilizing a semi-empirical coronal model, along with HUXt (Heliospheric Upwind eXtrapolation with time-dependence) and cone-CME models for the inner heliosphere, to forecast solar wind velocity and the arrival of CMEs associated with this event. The simulations were also carried out using Space Weather Adaptive SimulaTion (SWASTi) and a drag-based model (DBM) for this complex event of multiple CMEs. Predicted arrival times and velocities from these models are compared with actual observations at the Sun-Earth L1 point. These simulations reveal that three
coronal mass ejections (CMEs) reached Earth nearly simultaneously, resulting in the extreme space weather event, followed by the arrival of a few more eruptions. The simulations accurately predicted arrival times with a discrepancy of approximately 5 hours or less for these CMEs. Further, the ensemble study of DBM shows the sensitivity of the CME arrival time to the background solar wind speed and drag parameters. All three models have done fairly well in reproducing the arrival time closely to the actual observation of the CMEs responsible for the extreme geomagnetic storm of May 10, 2024. These rare solar storms offered a unique opportunity to thoroughly evaluate and validate our advanced models for predicting their arrival on the Earth.

\end{abstract}


\section{Introduction}

Extreme space weather events, driven by intense solar activity, can be initiated by powerful solar flares and coronal mass ejections (CMEs) at the Sun. The arrival of these transients from the Sun causes geomagnetic storms in Earth's near-space environment, that have the potential to severely disrupt our technological systems. A notable historical example is the Carrington Event of 1859, which caused widespread telegraph failures and auroras at unusually low latitudes. Another case is the Halloween Storms which occurred during solar cycle 23. These were a series of intense solar storms between late October and early November 2003, driven by a series of coronal mass ejections (CMEs). The Halloween storms caused widespread disruptions, including satellite anomalies, GPS navigation issues, and power system disturbances across various regions, exposing the vulnerability of modern technological infrastructure to such extreme space weather events. The intense storms also caused auroras to be visible at much lower latitudes than usual. \\

In the current solar cycle (SC-25), a powerful solar storm impacted Earth in early May 2024. This event was triggered by the intense activity from the active region AR13664 of the Sun. This region unleashed a series of X-class flares and coronal mass ejections (CMEs) during 8-9 May 2024, that were directed toward Earth. The resulting geomagnetic storm was very intense reaching extreme (NOAA G5) levels  (Minimum Dst index $\sim$-412 nT). This is the biggest Geomagnetic storm since 2003 in terms of its strength, and the active region on the Sun was as big as the Carrington event.  This series of events is now being called the \lq Gannon Superstorm of 2024 \rq in honor of the space weather physicist Jenn Gannon \citep{Yamazaki2024}. Such extreme events underscore the importance of space weather forecasting, not only to the research community but also to the governments, space sector, and industry stakeholders \citep{SCHRIJVER2015}.\\

The models that predict the arrival of these events generally use Solar surface magnetic field observations as their basic input and use either empirical or semi-empirical methods,  or magnetohydrodynamics (MHD) based simulations to arrive at solar wind parameters at different points in the heliosphere. Almost all models use a coupled two-domain procedure to arrive at solar wind parameters near the Earth and other planets.  In this study, we use the WSA approach in the coronal domain and the HUXt (heliospheric upwind extrapolation with time-dependence) \citep{OWENS2020, BARNARD2022} with CONE-CME model in the inner heliospheric domain, to arrive at the temporal variation of solar wind velocity for the period 9-15 May 2024. The predicted arrival times and velocities are compared with the ensemble drag-based model as well as actual observations at the Sun-Earth L1 point.

\section {Model Details}
\subsection{HUXt based Solar Transient ARrival (STAR) framework} \label{sec: STAR_intro}
This numerical framework for predicting the solar wind velocity and arrival of CMEs is based on established schemes that use a semi-empirical coronal model along with the HUXt, \citep{OWENS2020,BARNARD2022} and cone-CME models for the inner heliosphere.  Fig.~\ref{fig:flowchart} summarizes the details of the STAR module. Unlike MHD models, this module is capable of simulating the solar wind velocity alone and not the other properties like density, magnetic field, and temperature. However, this approach is computationally simple and inexpensive to analyze the arrival of transient events, since it is based on the HUXt code. The basic input to the coronal domain is the synoptic maps of magnetograms, provided by the Global Oscillation Network Group  (GONG) or HMI maps.
The Potential-Field Source-Surface (PFSS)  model solves the Laplace equation from the solar surface to a source surface (at 2.5 R$_s$,  where R$_s$ is the solar radius) from where the ﬁeld is assumed to be radial. We have used the PFSSPY package \citep{STANSBY2020} for this calculation, which is solved on a rectilinear grid that is equally spaced in $ln(r)$, cos$\theta$, and $\phi$ in spherical coordinates. A grid resolution of 100 $\times$ 181 $\times$ 361 is used to solve for ﬁeld lines from the solar surface to the source surface which is at a radius of 2.5 R$_s$ in our case. The field lines are then traced to get the regions of closed and open ﬁeld lines and the coronal hole boundaries. The expansion factor of open field lines ($f$) is then obtained  by, 
\begin{equation}
    f=\frac{R_s}{R_{ss}}\frac{B_r(R_s,\theta, \phi)}{B_{r}(R_{ss},\theta, \phi)}.
	\label{eq:exp factor}
\end{equation}
This equation is the same as that given in the previous works (e.g. \citep{POMOELL2018, MAYANK2022}, which deﬁnes the ﬂux tube expansion factor using only the radial magnetic ﬁeld component.  In addition to this factor, the great circle angular distance $d$ from the foot point of each open ﬁeld line to the nearest coronal hole boundary is also computed.
These two factors are used in the following empirical relation for solar wind velocities characterizing the ambient solar wind at 21.5$R_s$ for a given ﬂux tube.  We have adopted the following equation:
\begin{equation}
    v(f,d)=v_0+\frac{v_1}{(1+f)^\alpha}[(1-0.8 exp{(-d/w)^\beta)})^3]
	\label{eq:WSA}
\end{equation}
where $v_0$ = 250 km/s, $v_1$ = 650 km/s, $\alpha$ =2/9, $w$ = median of $d$ , and $\beta$ = 1.25.\\
It may be noted that several different forms for calculating the solar wind speed as a function of  $f$ and $d$ are available in the literature (e.g., \citet{ARGE2000, RILEY2001, MCGREGOR2011, POMOELL2018, MAYANK2022}).

The equation we used is the same as that used by \citet{MAYANK2022} for their MHD model, in which $w$ is taken as the median of $d$. In several other works, $w$ is taken as a constant value, for e.g., $w$ is 0.02 radians in \citet{POMOELL2018}. However, a difference from  \citet{MAYANK2022} is that they have used $v_0 = 240 km/s, v_1 = 725 km/s$ for their HUX  (not the time-dependent version) run, but we have found that for HUXt, $v_0 = 250 km/s, v_1 = 650 km/s$ is appropriate.

The equation (2)  provides the inner boundary conditions to the HUXt, which
 is based on incompressible hydrodynamics, which takes a reduced-physics approach to calculate only the solar wind velocity, employing approximations to greatly reduce the complexity of the MHD momentum equation. In this computation, the solar wind is assumed to be purely radial and is modeled by Burger's equation \citep{BARNARD2022}. Since HUXt maintains explicit time dependence, structures such as CMEs can be incorporated \citep{BARNARD2022}. To do this, the STAR code employs the cone CME model \citep{ODSTRCIL1999} which treats the CMEs as hydrodynamic clouds and are introduced as a time-dependent boundary condition at the inner radial boundary at 21.5$R_s$. The cone model parameters are obtained from the CCMC Space Weather Database Of Notiﬁcations, Knowledge, Information (DONKI) database.\\
The PFSSPY \citep{STANSBY2020} code is obtained from \url{https://pfsspy.readthedocs.io/en/latest/installing.html}. The HUXt code \citep{OWENS2024_software}  is downloaded from \url{https://zenodo.org/records/10842659} and utilized for further analysis.
\subsection{Space Weather Adaptive Simulation framework (SWASTi)}

We have also used the Space Weather Adaptive Simulation framework \citep[SWASTi;][]{MAYANK2022, Mayank2024} to study this highly geo-effective event. SWASTi is a physics-based modular framework that solves the magnetohydrodynamic (MHD) equations to simulate the solar wind and CMEs in the inner heliosphere. In this study, the MHD domain extends radially from 0.1 AU to 2.1 AU. The simulation employs a synoptic magnetogram as the observational input for the solar wind model, with the initial solar wind speed ($V_r$) at 0.1 AU determined using a modified WSA speed relation (see Equation \ref{eq:WSA}), and other initial properties calculated using the following empirical relations based on fast solar wind characteristics:

    \begin{align}
        \label{eq9} n  &=  n_0 \, \bigg(\frac{V_{fsw}}{V_r}\bigg)^{\,2} \\
        \label{eq10} B_r  &=  \text{sgn} (B_{corona})\,B_0\,\bigg(\frac{V_r}{V_{fsw}}\bigg) \\
        \label{eq11} B_{\phi}  &=  -\,B_{r}\,\sin{\theta}\,\bigg(\frac{V_{rot}}{V_r}\bigg)
   \end{align}

where $n$ is the plasma number density, $B_r$ and $B_{\phi}$ are the radial and azimuthal components of the magnetic field, $sgn(B_{corona})$ is the polarity of the extrapolated coronal magnetic field, and $V_{rot}$ is the rotational speed of the inner boundary corresponding to the time span of the magnetogram. Here, $n_0$ and $B_0$ refer to the number density and magnetic field values of the fast solar wind with speed $V_{fsw}$. In this work, the default values of these parameters are taken, which are: $n_0$ = 200 $cm^{-3}$, $B_0$ = 300 nT, and $V_{fsw}$ = 650 km/s. The initial thermal pressure (at 0.1 AU) is kept constant at 6.0 nPa and the meridional \& azimuthal components of velocity ($V_\theta$ and $V_\phi$) are assumed to be zero initially.

Within the SWASTi framework, there are two distinct CME models: the elliptic cone and flux rope CME models. The cone model features a simple non-magnetic geometry, whereas the flux rope model incorporates a complex 3D magnetic field configuration \citep{Mayank2024}. To ensure consistency between the MHD simulation and the results of STAR and DBM-based CMEs in this work, we used the cone model with the same initial properties described in Table \ref{tab:cme_data}. The half-width and half-height of the CME are set equal to the half-angle. Additionally, the CME properties at 21.5 R$_s$, including arrival time and speed, as well as the eruption coordinates on the solar surface, are consistent with those listed in Table \ref{tab:cme_data}. For simplicity, we have assumed a constant temperature of 0.8 MK and a density of 2$\times$10$^{-18}$ kg/m$^3$ for all CMEs.

\subsection{Drag Based Model (DBM) for CME propagation and arrival time}
The drag-based models rely on MHD drag, which is caused by the emission of MHD waves in the collisionless solar wind, rather than kinetic drag in a fluid \citep{Cargill_1996}. This drag modulates the speed of coronal mass ejections (CMEs) to that of the surrounding solar wind. In interplanetary space, CMEs are influenced by the Lorentz force ($F_{L}$), gravity ($F_{G}$), and MHD aerodynamic drag ($F_{D}$).
Thus the net force acting on the CME can be expressed as,
\begin{equation}
    F = F_{L}-F_{G} + F_{D}
\end{equation}

 The analytical drag-based model (DBM), are commonly used model for the heliospheric propagation of CMEs, and is valued for its simplicity and quick calculations. DBM considers CME propagation as a momentum exchange with the ambient solar wind.  However, the DBM is mainly driven by magnetohydrodynamic (MHD) drag, which aligns the CME speed with the ambient solar wind. 


The following equation from \cite{Cargill_2004} represents the propagation of the CMEs or shock in the background solar wind. 
\begin{equation}
    a(t)=-\gamma \big(V_{CME}(t)-V_{SW}\big)|V_{CME}(t)-V_{SW}|
\end{equation}
where $a(t)=\frac{d^2 R(t)}{dt^2}$ represents the CME acceleration, $V_{CME}(t)$ represents the CME speed, $R(t)$ represents the heliospheric distance, $\gamma$ represents the drag parameter, which describes the rate of change of CME speed and is assumed to be constant, and $V_{SW}$ represents the solar wind speed, also assumed to be constant. Further, the advanced drag-based model (ADBM) accounts for the shape of CMEs and solves the equation of motion for the CME geometry. The geometry could be approximated by a flat front, semi-circular, or circular \citep{Schwenn_2005}. The propagation of CME can be determined by analyzing the evolution of two types of 2D-cone geometry: (a) self-similar cone evolution and (b) flattening cone evolution \citep{Dumbovi__2021}. The DBM has been compared with other existing models of CME propagation and found to be doing fairly well \citep{Vr_nak_2014}. The DBM takes the following inputs: CME speed ($V_{CME}$) at certain distance, half-width of CME($\theta_{CME}$), and propagation direction (longitude, $\phi_{CME}$), the solar wind radial speed ($V_{SW}$), the drag parameter ($\gamma$), target direction (longitude, $\phi_{T}$). The codes of the Advanced DBM (ADBM) model open-source codes available at (\url{https://zenodo.org/record/5038648}) are adapted for the study.  The determination of initial parameters like background solar wind speed ($V_{sw}$) and drag parameter ($\gamma$) is always the open issue \citep{Dumbovi__2021}. Various approaches have been adopted based on physics-based/empirical models and in situ observations.  It has been observed $V_{sw}=450$ km/sec and $\gamma= 0.2 \times 10^{-7}$ are optimum parameters \citep{Vr_nak_2012,Vr_nak_2014}, which generally work for high or low solar activity.  Here we used these values as mean values and assumed their Gaussian distribution to study their impact on CME arrival time and velocity at Earth. 


As mentioned, the DBM models CME propagation as a momentum exchange between the CME and the ambient solar wind, assuming an ambient solar wind density that falls off as \( \frac{1}{r^2} \). This results in the ICME cross-section increasing as \( A \propto r^2 \) with constant mass and \( \gamma \). However, other studies show the CME cross-section increases as \( A \propto r^{1.6} \), which is slower than the DBM assumption \citep{Bothmer_1997}. The DBM assumes isotropic, constant-speed solar wind, oversimplifying since ICME propagation can alter the solar wind structure. It also ignores CME-CME interactions, where a fast CME can accelerate a slower one launched earlier. However, this is computationally very fast and provides realistic predictions. The model has been compared with the WSA-ENLIL+CONE model and found to be reasonably comparable in the prediction of CME arrival time \citep{Vr_nak_2014}. 


\section{Data}
As mentioned, the STAR module is capable of generating the inner boundary conditions for HUXt, by using GONG synoptic, GONG-ADAPT (Air Force Data Assimilative Photospheric Flux Transport, \citet{HICKMANN2015}), and HMI Maps.  For the present study, we used both GONG-synoptic and  GONG-ADAPT as inputs, for solving the Potential Field Source Surface (PFSS) coronal model. For providing the synoptic magnetograms required by the PFSS
model, both the hourly updated standard synoptic magnetograms (GONG-synoptic) maps and the  
ADAPT maps are used. For this study, the hourly updated integral Carrington rotation (CR) synoptic
maps are downloaded from \url{https://gong.nso.edu/data/magmap/QR/bqj}, for CR2284. We have used the GONG maps generated on 9 May 2024 and 10 May 2024 in different simulations. These maps contain information of the photospheric magnetic field corresponding to the period from 2024 April 17 15:28 UTC to 2024-05-14 21:12 UTC. The six, two-hourly  ADAPT maps obtained on 10 May 2024 (00:00-10:00 UTC) downloaded from \url{https://gong.nso.edu/adapt/maps/gong/2024}, are also used for simulations. 

The CME parameters are taken from the DONKI website (\url{https://kauai.ccmc.gsfc.nasa.gov/DONKI/search/}). For the given time period, the DONKI database lists many CMEs, of which some are slow (less than 500 km/s) or narrow (half-width less than 35$^{\circ}$) or not directed towards Earth (source longitude not within $\pm60^{\circ}$ HEEQ), are excluded and the remaining 6 CMEs erupted on 8-9 May 2024 are selected for modeling. The selected CMEs originated from the AR13664 of the Sun. The parameters of the six selected CMEs are provided in Table 1.  In these, the CMEs 1-5 are used in all STAR, DBM and SWASTi simulations whereas CME-6 is only used in certain cases, as explained later. The Interplanetary Magnetic Field (IMF), solar wind speed, density, and dynamic pressure,  measurements near Sun-Earth L1 point as well as the Sym-H (representing the ring current) variation
at Earth are obtained from the NASA Space Physics Data Facility (SPDF) OMNIWeb data center, with 5-minute temporal resolution (\url{https://omniweb.gsfc.nasa.gov/}). The actual CME shock arrival times are taken from \url{https://kauai.ccmc.gsfc.nasa.gov/CMEscoreboard/}. Additionally, the CME arrival time corresponding to MHD simulation using the SWASTi framework is also presented in this study. For the MHD simulations, we have used the GONG-ADAPT magnetogram.

\begin{table*}[!h]
    \centering
    \resizebox{\textwidth}{!}{%
    \begin{tabular}{|c|c|c|c|c|c|c|c|}
        \hline
        \textbf{CME ID} & \textbf{CME Initiation Time} & \textbf{Time at 21.5 \(R_s\)} & \textbf{CME Speed (km/s)} & \textbf{Longitude (°)} & \textbf{Latitude (°)} & \textbf{Half Angle (°)} \\
        \hline
        CME-1 & 2024-05-08T05:36 & 2024-05-08T09:30 & 870.0 & 9.0 & -7.0 & 43.0  \\
        \hline
        CME-2 & 2024-05-08T12:24 & 2024-05-08T16:58 & 776.0 & 3.0 & -5.0 & 45.0  \\
        \hline
        CME-3 & 2024-05-08T19:12 & 2024-05-08T23:27 & 828.0 & -23.0 & 3.0 & 38.0  \\
        \hline
        CME-4 & 2024-05-09T18:23 & 2024-05-09T20:27 & 1236.0 & 12.0 & -5.0 & 45.0   \\
        \hline
        CME-5 & 2024-05-10T07:12 & 2024-05-10T10:35 & 1018.0 & 31.0 & -2.0 & 41.0   \\
        \hline
      CME-6 & 2024-05-11T01:36 & 2024-05-11T04:22 & 1263.0 & 51.0 & 0.0 & 51.0 \\
         \hline
    \end{tabular}}
    \caption{Summary of the Input CME parameters used for different simulations.}
    \label{tab:cme_data}
\end{table*}

 \section{Results}
 Fig.~\ref{fig:bkg} shows the solar wind parameters and the geomagnetic SYM-H index for the period 8-14 May 2024. The parameters shown here are solar wind velocity (V), density (Den.), Dynamic Pressure (Dyn.P), Temperature (T),  Total interplanetary magnetic field (IMF) and its components, as well as the geomagnetic SYM-H index in panels (a-e) respectively. It can be seen that the arrival of the first shock is clearly seen around 16:45 UTC on 10 May 2014, with an enhancement in solar wind speed, density as well as magnetic field. A possible flux rope signature can be observed around $\sim$ 21:30 UTC which corresponds to a second, more intense amplification of magnetic field components with $\lvert B\lvert$  reaching a maximum of $\sim$ 75 nT. This flux rope feature, which extends to 09:30 on 11 May, there are three noticeable decreases in density which may correspond to three CMEs that may have merged into the signature.  The southward $B_z$ component became almost -50 nT in the early hours of 11 May 2024.  The enhanced solar wind density and speed (750-950 km/s) prevailed between 11 and 12 May. The maximum SymH value of $\sim$ -412 nT was observed around 03:00 UTC on 11 May. Though the major geomagnetic field enhancements lasted only up to 12 May, the velocity continued to remain high on 13 May as well, with another enhancement of density with lesser magnitude compared to the previous days.

Fig.~\ref{fig:gong} shows the results of the STAR simulations for 9-15 May 2024, with GONG maps as input, along with the solar wind speed observations at L1. Fig.~\ref{fig:gong}a shows the simulation done with the GONG map obtained on 09 May 2024 at 23:14 UTC. The output of the simulation is shifted backward by 3.5 hours to match with the first CME arrival as shown in the figure.
The motivation behind applying the time shift to the simulations is to compare the observed and simulated CME-related enhancements in speeds and their temporal variation in a scenario had we predicted the arrivals correctly. The time shifting also allows for quantification of the error in CME arrival time for the best-fit result.

Five CMEs are used for this simulation, and it is seen that the first three CMEs arrive nearly simultaneously to generate the first peak in velocity, followed by the subsequent arrivals of the other two CMEs. It can be seen that solar wind speed of the first merged CME arrival is also higher than that observed at L1. The subsequent peaks are more or less comparable with observations but shifted in terms of structure and arrival times. The overall RMSE is 100.6 km/s, with time-shifted simulation output and observations. \\
Fig.~\ref{fig:gong}b shows the simulation done with the GONG map obtained on 10 May 2024 at 08: 04UTC. Here we can see that the magnitude of the initial background velocity as well as the velocity peaks due to the arrival of the CMEs match better with the observations. However, the time series is shifted by 8 hours to match with the first CME arrival. The lower background velocity in this case might have allowed only a slower propagation for the CMEs in the inner heliosphere domain of the model. The overall RMSE is 112.2 km/s, with a 8-hour shifted simulation output and observations.

An interesting point to note here is that the simulated solar wind velocity steadily decreased after 12 May and the enhanced solar wind observed on 13 May is not reproduced. Therefore, we have incorporated one more CME  (CME-6 in table-1) into the model and  Fig.~\ref{fig:gong}c shows the output. This has allowed the model to somewhat capture the enhancement in solar wind speed on 13 May, and the RSME between time-shifted model output and observations is reduced to 94 km/s.  These simulations convincingly reveal these aspects: (a) three CMEs (CME1, CME2, and CME3 in table-1) arrive at L1 almost simultaneously and cause the first enhancement in the solar wind parameters. The CME3 is the fastest and takes the least time to arrive at Earth. The CME1 takes the longest time to arrive at Earth. (b) two more faster CMEs (CME4 and CME5 in table-1) erupted on 9 and 10 May and arrived on Earth subsequently on the next days and (c) the passage of yet another CME (CME6 in table-1) is the probable cause of the continued enhancement in solar wind velocity observed on 13 May 2024.

Fig.~\ref{fig:adapt} shows the average of the 6 simulations using the 6 ADAPT maps obtained on 10 May 2024, at least $\sim$ 6 hours before the arrival of the event (00 UTC, 02 UTC, 04 UTC, 06 UTC, 08 UTC and 10 UTC) and 6 CMEs as inputs to the model. Here time-shift applied to match with the observations is 5 hours, and arrival times for each simulation differ only by minutes. The RMSE for the average of the time-shifted ADAPT simulations and observations is  91.459 km/s. 
Fig.~\ref{fig:huxmaps} shows the snapshots of the radial maps from these simulations.  Fig.~\ref{fig:huxmaps}a shows the condition before the CME arrival at L1, and just after the launch of the first CME (on 08 May 05:36 UTC at 21.5R$\_s$) into the model. Fig.~\ref{fig:huxmaps}b shows the snapshot close to the arrival of the three CMEs (merged) at L1. The CME4 and CME5 (launched on 09 May 18:23 UTC and 10 May 07:12 UTC respectively) are also seen in the inner heliosphere. Fig.~\ref{fig:huxmaps}c shows the snapshot close to the arrival of the CME4, whereas the CME5 is still behind. We can see that the CME1, CME2, and CME3 arrived on Earth nearly simultaneously, whereas the CME4 and CME5 arrived at distinctly different times. Finally, Fig.~\ref{fig:huxmaps}d shows the glancing arrival of CME6 on 13 May, which could only cause velocity and density enhancements without change in IMF magnitude or direction.

All six CMEs were also simulated using the cone model of the SWASTi framework. Fig. \ref{fig:SWASTi_plot} shows the results of this simulation. The top two panels illustrate the speed and scaled density profiles in the equatorial plane on 12 May 2024 at UTC 05:16. Four CME shock fronts are visible in these 2D plots. The first front represents the merged structure of the initial three CMEs ( CME1, CME2, and CME3), which had already crossed the Earth's location by that time. These first three CMEs interacted with each other, causing their shocks to merge into a single front before reaching 1 AU. This merging is also evident in the in-situ speed profile shown in the bottom panel of Fig. \ref{fig:SWASTi_plot}. The second CME front, corresponding to CME4, is also visible in subplots \ref{fig:SWASTi_plot}a and \ref{fig:SWASTi_plot}b subplots, and has also passed the Earth's location. At the time of the snapshot, CME5 is passing through this location, while CME6 is still behind.

When comparing the simulation results with ACE spacecraft observations, we found that the merged structure of CME1, CME2, and CME3 arrived approximately 0.97 hours later than the observed arrival time. Additionally, the apex speed of the CMEs is almost the same in both the ACE (blue line) and SWASTi (red line) plots. The speed of the subsequent CME (CME4) front also shows a similar value in both in-situ plots. However, there is a significant difference of approximately +11.2 hours in the simulated arrival time of CME4 compared to the DONKI arrival time (see Fig. \ref{fig:SWASTi_plot}c). It is important to note that the uncertainty regarding the initial density of the CME can significantly affect its arrival time. To demonstrate this, we conducted an additional simulation by reducing the density of CME4 to 0.5$\times$10$^{-18}$ kg/m$^3$.

The in-situ plot of this additional SWASTi simulation (SWASTi - tuned) is represented by the orange line in Fig. \ref{fig:SWASTi_plot}c. A time difference of about 5.9 hours can be observed between the shock arrival times of the default and tuned versions of CME4. Additionally, there is a difference of roughly 150 km/s in the apex speeds of CME4. Interestingly, the change in the initial density of CME4 not only affects the arrival properties of CME4 itself but also impacts those of CME5 and CME6. This showcases the importance of estimating initial density of CMEs to forecast their arrival time and properties accurately. This point is further highlighted by the fact that the overall RMSE of the default (SWASTi - 1) run (from 9 to 15 May 2024) is approximately 120 km/s, while that of the tuned (SWASTi- 2) run is about 100 km/s, reflecting a difference of 16.67\% in overall accuracy.

As mentioned earlier, we have used DBM, and flattening top geometry is used where each element of CME-front propagated independently and tracked in the heliosphere. Fig.~\ref{fig:geometry_dbm} shows the heliospheric snapshot of all five CMEs when CME-1 reached the Earth. Note that individual CMEs propagated based on the drag model and CME-CME interaction are not accounted for here. The DBM shows that all these 5 CMEs directly or obliquely intersected the Earth.  Next Fig.~\ref{fig:RT} shows how the velocity of each CME evolved during their radial-outward propagation in the interplanetary space. All CMEs started with high initial velocity at 21.5 $R_{s}$ and naturally, their propagation velocity decreases as they move radially outward into the heliosphere.  In this figure, the distance-time path (dashed line) of the front-apex of each CME is shown. The distance-time paths come very close to each other for most of the CMEs and CMEs are extended objects, indicating possible interaction between them before they reach the Earth. Based on the model we can estimate each CME's arrival time and velocity. However, the arrival time and velocity are subject to change depending on the uncertainty in the initial parameters of CME and background solar wind conditions. In reality, each measurement has some uncertainty, sometimes it is known, and sometimes it is unknown. Hence we tried to conduct a sensitivity analysis of two variables: background solar wind velocity (Vsw) and drag parameters ($\gamma$). These parameters are varied assuming Gaussian distributions (Fig.~\ref{fig:hist_input}) and then using these ensemble samples the arrival time, arrival velocity of CME, and transit time of CME for each pair of values, were estimated.  Fig.~\ref{fig:hist_input} shows the distribution of Vsw and $\gamma$ as these parameters represent background conditions for CME propagation.  The mean background solar wind velocity was considered to be $450km/sec$,  having $3\sigma=50km/sec$ whereas,  mean drag parameters ($\gamma$) was considered to be $2\times 10^{-8}$,  having $3\sigma=0.5\times 10^{-8}$.  Fig.~\ref{fig:hist_output} presents the distributions of CME arrival velocity, arrival time at Earth, and transit time from the Sun to Earth for each CME listed in  Table ~\ref{tab:cme_data}, obtained using the ensemble of inputs. The broad distribution of CME arrival velocity, arrival time, and transit time highlights the significant influence of the background solar wind speed and the drag parameter on the accuracy of forecasting CME arrivals at Earth \citep{Mayank2024, Kay2023}. The range of arrival velocities within the ensemble spans approximately $\sigma \sim 50$ km/s around the mean value and arrival times range $\sigma \sim 3 $ hr. Whereas the minimum to maximum values of arrival velocity spans about $\pm 100$ km/s, and the arrival time spans about $\pm 6$ hours from the respective mean values. This variability underscores the complexities involved in predicting CME behavior and the need for precise modeling of the solar wind and drag effects to improve forecast accuracy.

Table \ref{tab:cme_arrival} gives a summary of simulated CME arrival times and speeds, using both STAR, SWASTi code, and DBM, along with the observed CME shock arrival times. For the STAR simulations, the results using the GONG-ADAPT maps are given. It can be seen that in the actual case also, the first three CMEs arrived simultaneously, which was re-produced by STAR and SWASTi.  For the CME-4 and CME-5, the difference between the arrival times is less than 2 hours as modeled by the STAR. The shock related to the CME-6 is not observed near Earth. The STAR simulations also show that the CME shock front did not directly hit Earth. Whereas, DBM shows different arrival times of all these CMEs, with CME-1 reaching almost 1 hour earlier than the observed shock arrival time. DBM run for CME-2 and 3  shows arrival around 9 and 17 hours later than observed respectively.  For the CME-4 and CME-5, the difference between the arrival times is around 3-8 hours.   This indicates the STAR and SWASTi very closely predicted the arrival times of CMEs as compared to DBM. Note that STAR and SWASTi account for the CME-CME interaction whereas DBM does not account for this.  However, the arrival of CME-1 is very well predicted by all the studied models.

\begin{table*}
\hspace{-1cm}
   \resizebox{\textwidth}{!}{%
    \begin{tabular}{|c|c|c|c|c|c|c|c|}
        \hline
        \textbf{CME ID} & \textbf{DONKI Arrival time} & \textbf{Arrival:STAR} & \textbf{Speed:STAR} & \textbf{Arrival:DBM} & \textbf{Speed:DBM} & \textbf{Arrival:SWASTi} & \textbf{Speed:SWASTi} \\
        \hline
        CME-1 &2024-05-10T16:36 &-5.38  & 616 & +1.08 $\pm$ 3.0hr & 609.9  $\pm$ 48.3 & -0.97 hr & 693    \\
        \hline
        CME-2 &2024-05-10T16:36 &-5.38 & 616 & -9.48 $\pm$ 2.9hr& 590.1  $\pm$ 43.9 & -0.97 hr & 693   \\ 
        \hline
        CME-3 &2024-05-10T16:36 &-5.38 & 616 & -17.73 $\pm$ 2.9hr& 580.7  $\pm$ 42.2 & -0.97 hr & 693   \\
        \hline
        CME-4 &2024-05-11T20:30 &-1.53 &745 & +3.40 $\pm$ 3.2hr & 675.3  $\pm$ 63.6 & +5.30 hr & 968    \\
        \hline
        CME-5 &2024-05-12T08:55 & -0.43  &824 & -8.72 $\pm$ 3.0hr& 605.8  $\pm$ 47.9 & +5.85 hr & 975   \\
        \hline
    \end{tabular}}
    \caption{Summary of simulated CME arrival times and speeds(km/s) at the Earth. For the STAR simulations, the results using the GONG-ADAPT map is given. For the DBM simulations, the mean speed and arrival times are given. The CME arrival time and speed corresponding to SWASTi - 2 simulation are also shown. The actual CME shock arrival times are taken from \url{https://kauai.ccmc.gsfc.nasa.gov/CMEscoreboard/}. The difference is taken from the observations with the model, hence whenever the model simulated arrivals are late, they are negative. }
    \label{tab:cme_arrival}
\end{table*}

\section{Discussion and Conclusions}

The HUXt code has been developed in such a way that the boundary conditions can be accepted from a wide range of coronal models, including potential field source surface-based models such as WSA \citep{ARGE2000} and MHD models such as Magnetohydrodynamic Algorithm outside a Sphere (MAS, \citet{RILEY2001}) and conditions derived from tomography such as CORTOM \citep{BUNTING2022}. However, in most of the previous works \citep{OWENS2020} the HUXt has been initiated with output from the MAS model. For example,\citet{OWENS2020} performed the HUXt analysis with steady-state HelioMAS solutions as input for a period of 578 Carrington Rotations. Based on several cases they showed that the estimated CME transit times were agreed to within four hours. In another recent study, the output from the  Burger Radial Variational Data Assimilation (BRaVDA) scheme is used to define the solar wind speed structure at the inner boundary of HUXt \citep{JAMES2023}. \citep{BARNARD2023} has used a sequential importance resampling (SIR) data assimilation scheme with the HUXt solar wind model and performed a set of theoretical experiments to show that SIR-HUXt can reduce the uncertainty on the CME arrival time and speed estimates, dropping by up to 69$\%$ and 63$\%$ for an observer at the L5 Lagrange point.  However, in this work, no case studies of real CMEs in structured solar wind are shown. 

In the present STAR framework, the inner boundary conditions for the HUXt module are provided from the semi-empirical WSA relation obtained from PFSSPY making use of GONG maps, and CMEs are introduced via the cone CME parameterization to the model inner boundary. These simulations show that for this extreme space weather event, the WSA + HUXt + cone CME-based simple simulations could predict the arrival times within a difference of $\sim$ 8 hours (CMEs 1--3) or less (in the case of CME4 and 5), which is comparable to the predictions of complex, computationally expensive MHD codes.  

Simulations using the cone model in the SWASTi framework revealed that the merged shock front of CME 1, 2, and 3 arrived 0.97 hours later than observed by ACE, while CME4's simulated arrival was delayed by 11.2 hours compared to DONKI predictions. Adjusting CME4's initial density in a secondary simulation improved arrival time accuracy by 5.9 hours and reduced speed discrepancies, highlighting the critical role of accurate density estimation in forecasting CME properties.

Further, we compared these simulations by the STAR and SWASTi models to DBM model outputs: arrival time and ICME velocity at the Earth. It is observed that all models predicted the arrival time of the CMEs reasonably well. We also carried out the ensemble study for DBM for varying background solar wind conditions and drag parameters. It has been observed that the arrival time and velocity of CME are sensitive to these background conditions which are generally unknown, as also shown by other model simulations (e.g. \citet{Mayank2024, Dumbovi__2018}). Even though we have not considered the uncertainty in the CME parameters, which definitely will affect arrival predictions, the DBM is still reasonably accurate in forecasting CME arrival time and velocity. 
The models for predicting the arrival of space weather events are generally able to get the  CME shock arrival times to within $\pm$ 10 hours, but with standard deviations often exceeding 20 hours \citep{Riley2018}.

The case study of a recent extreme space weather event provided a unique opportunity to evaluate the performance of the models for ICME arrival. For this event, the predictive simulations expected the arrival of ICME   on late 10 May or Early 11 May, however the event arrived earlier compared to the predicted arrival times of several models(\url{https://kauai.ccmc.gsfc.nasa.gov/CMEscoreboard/}), including the present STAR, SWASTi, and DBM model. This implies that there is still improvement needed in models to match the observations, and future work will strive to improve the forecast of the arrival time of such extreme space weather events. 

\newpage
\section*{Acknowledgements}
The work is supported by the Indian Space Research Organization. Prateek Mayank
gratefully acknowledges the support provided by the Prime Minister's Research Fellowship. Indu Venugopal acknowledges the financial assistance provided by ISRO through a research fellowship. This work utilizes data produced collaboratively between the Air Force Research Laboratory (AFRL) \& the National Solar Observatory (NSO). The ADAPT model development is supported by AFRL. The input data utilized by ADAPT is obtained by NSO/NISP (NSO Integrated Synoptic Program). The GONG synoptic maps are taken from \url{https://gong.nso.edu/data/magmap/}. NSO is operated by the Association of Universities for Research in Astronomy (AURA), Inc., under a cooperative agreement with the National Science Foundation (NSF). The solar wind data is taken from the OMNIweb database (\url{https://omniweb.gsfc.nasa.gov/}). We thank the OMNIweb team for the data. This research made use of Astropy (\url{http://www.astropy.org}),
a community-developed core Python package for Astronomy, 
version 4.0.0 of the SunPy {\url{https://sunpy.org/}and PFSSPY open source software packages. The HUXt code downloaded from \url{https://github.com/University-of-Reading-Space-Science/HUXt/tree/v.4.1.1, https://zenodo.org/records/10842659} and the Advanced DBM model codes available at (\url{https://zenodo.org/record/5038648}) are used in this work.  We express our sincere thanks to the developers for providing these codes.

\bibliography{Smitha_new_bibliography}{}

\begin{thebibliography}{}
\expandafter\ifx\csname natexlab\endcsname\relax\def\natexlab#1{#1}\fi
\providecommand{\url}[1]{\href{#1}{#1}}
\providecommand{\dodoi}[1]{doi:~\href{http://doi.org/#1}{\nolinkurl{#1}}}
\providecommand{\doeprint}[1]{\href{http://ascl.net/#1}{\nolinkurl{http://ascl.net/#1}}}
\providecommand{\doarXiv}[1]{\href{https://arxiv.org/abs/#1}{\nolinkurl{https://arxiv.org/abs/#1}}}

\bibitem[{Arge \& Pizzo(2000)}]{ARGE2000}
Arge, C.~N., \& Pizzo, V.~J. 2000, Journal of Geophysical Research: Space Physics, 105, 10465, \dodoi{https://doi.org/10.1029/1999JA000262}

\bibitem[{Barnard \& Owens(2022)}]{BARNARD2022}
Barnard, L., \& Owens, M. 2022, Frontiers in Physics, 10, \dodoi{10.3389/fphy.2022.1005621}

\bibitem[{Barnard {et~al.}(2023)Barnard, Owens, Scott, Lang, \& Lockwood}]{BARNARD2023}
Barnard, L., Owens, M., Scott, C., Lang, M., \& Lockwood, M. 2023, Space Weather, 21, e2023SW003487, \dodoi{https://doi.org/10.1029/2023SW003487}

\bibitem[{Bothmer \& Schwenn(1997)}]{Bothmer_1997}
Bothmer, V., \& Schwenn, R. 1997, Annales Geophysicae, 16, 1, \dodoi{10.1007/s005850050575}

\bibitem[{{Bunting} \& {Morgan}(2022)}]{BUNTING2022}
{Bunting}, K.~A., \& {Morgan}, H. 2022, Journal of Space Weather and Space Climate, 12, 30, \dodoi{10.1051/swsc/2022026}

\bibitem[{Cargill(2004)}]{Cargill_2004}
Cargill, P.~J. 2004, Solar Physics, 221, 135–149, \dodoi{10.1023/b:sola.0000033366.10725.a2}

\bibitem[{Cargill {et~al.}(1996)Cargill, Chen, Spicer, \& Zalesak}]{Cargill_1996}
Cargill, P.~J., Chen, J., Spicer, D.~S., \& Zalesak, S.~T. 1996, Journal of Geophysical Research: Space Physics, 101, 4855–4870, \dodoi{10.1029/95ja03769}

\bibitem[{Dumbović {et~al.}(2021)Dumbović, Čalogović, Martinić, Vršnak, Sudar, Temmer, \& Veronig}]{Dumbovi__2021}
Dumbović, M., Čalogović, J., Martinić, K., {et~al.} 2021, Frontiers in Astronomy and Space Sciences, 8, \dodoi{10.3389/fspas.2021.639986}

\bibitem[{Dumbović {et~al.}(2018)Dumbović, Čalogović, Vršnak, Temmer, Mays, Veronig, \& Piantschitsch}]{Dumbovi__2018}
Dumbović, M., Čalogović, J., Vršnak, B., {et~al.} 2018, The Astrophysical Journal, 854, 180, \dodoi{10.3847/1538-4357/aaaa66}

\bibitem[{{Hickmann} {et~al.}(2015){Hickmann}, {Godinez}, {Henney}, \& {Arge}}]{HICKMANN2015}
{Hickmann}, K.~S., {Godinez}, H.~C., {Henney}, C.~J., \& {Arge}, C.~N. 2015, \solphys, 290, 1105, \dodoi{10.1007/s11207-015-0666-3}

\bibitem[{James {et~al.}(2023)James, Scott, Barnard, Owens, Lang, \& Jones}]{JAMES2023}
James, L.~A., Scott, C.~J., Barnard, L.~A., {et~al.} 2023, Space Weather, 21, e2022SW003289, \dodoi{https://doi.org/10.1029/2022SW003289}

\bibitem[{Kay {et~al.}(2023)Kay, Nieves-Chinchilla, Hofmeister, Palmerio, \& Ledvina}]{Kay2023}
Kay, C., Nieves-Chinchilla, T., Hofmeister, S.~J., Palmerio, E., \& Ledvina, V.~E. 2023, Space Weather, 21, e2023SW003647, \dodoi{https://doi.org/10.1029/2023SW003647}

\bibitem[{Mayank {et~al.}(2022)Mayank, Vaidya, \& Chakrabarty}]{MAYANK2022}
Mayank, P., Vaidya, B., \& Chakrabarty, D. 2022, The Astrophysical Journal Supplement Series, 262, 23, \dodoi{10.3847/1538-4365/ac8551}

\bibitem[{Mayank {et~al.}(2023)Mayank, Vaidya, Mishra, \& Chakrabarty}]{Mayank2024}
Mayank, P., Vaidya, B., Mishra, W., \& Chakrabarty, D. 2023, The Astrophysical Journal Supplement Series, 270, 10, \dodoi{10.3847/1538-4365/ad08c7}

\bibitem[{McGregor {et~al.}(2011)McGregor, Hughes, Arge, Owens, \& Odstrcil}]{MCGREGOR2011}
McGregor, S.~L., Hughes, W.~J., Arge, C.~N., Owens, M.~J., \& Odstrcil, D. 2011, Journal of Geophysical Research: Space Physics, 116, \dodoi{https://doi.org/10.1029/2010JA015881}

\bibitem[{Odstrcil \& Pizzo(1999)}]{ODSTRCIL1999}
Odstrcil, D., \& Pizzo, V.~J. 1999, Journal of Geophysical Research: Space Physics, 104, 28225, \dodoi{https://doi.org/10.1029/1999JA900319}

\bibitem[{Owens \& Barnard(2024)}]{OWENS2024_software}
Owens, M., \& Barnard, L. 2024, {University-of-Reading-Space-Science/HUXt: Minor bug fixes and updated environment}, v.4.1.1,  Zenodo, \dodoi{10.5281/zenodo.10842659}

\bibitem[{{Owens} {et~al.}(2020){Owens}, {Lang}, {Barnard}, {Riley}, {Ben-Nun}, {Scott}, {Lockwood}, {Reiss}, {Arge}, \& {Gonzi}}]{OWENS2020}
{Owens}, M., {Lang}, M., {Barnard}, L., {et~al.} 2020, \solphys, 295, 43, \dodoi{10.1007/s11207-020-01605-3}

\bibitem[{{Pomoell, Jens} \& {Poedts, S.}(2018)}]{POMOELL2018}
{Pomoell, Jens}, \& {Poedts, S.} 2018, J. Space Weather Space Clim., 8, A35, \dodoi{10.1051/swsc/2018020}

\bibitem[{Riley {et~al.}(2001)Riley, Linker, \& Mikić}]{RILEY2001}
Riley, P., Linker, J.~A., \& Mikić, Z. 2001, Journal of Geophysical Research: Space Physics, 106, 15889, \dodoi{https://doi.org/10.1029/2000JA000121}

\bibitem[{Riley {et~al.}(2018)Riley, Mays, Andries, Amerstorfer, Biesecker, Delouille, Dumbović, Feng, Henley, Linker, Möstl, Nuñez, Pizzo, Temmer, Tobiska, Verbeke, West, \& Zhao}]{Riley2018}
Riley, P., Mays, M.~L., Andries, J., {et~al.} 2018, Space Weather, 16, 1245, \dodoi{https://doi.org/10.1029/2018SW001962}

\bibitem[{Schrijver {et~al.}(2015)Schrijver, Kauristie, Aylward, Denardini, Gibson, Glover, Gopalswamy, Grande, Hapgood, Heynderickx, Jakowski, Kalegaev, Lapenta, Linker, Liu, Mandrini, Mann, Nagatsuma, Nandy, Obara, {Paul O’Brien}, Onsager, Opgenoorth, Terkildsen, Valladares, \& Vilmer}]{SCHRIJVER2015}
Schrijver, C.~J., Kauristie, K., Aylward, A.~D., {et~al.} 2015, Advances in Space Research, 55, 2745, \dodoi{https://doi.org/10.1016/j.asr.2015.03.023}

\bibitem[{Schwenn {et~al.}(2005)Schwenn, Dal~Lago, Huttunen, \& Gonzalez}]{Schwenn_2005}
Schwenn, R., Dal~Lago, A., Huttunen, E., \& Gonzalez, W.~D. 2005, Annales Geophysicae, 23, 1033–1059, \dodoi{10.5194/angeo-23-1033-2005}

\bibitem[{Stansby {et~al.}(2020)Stansby, Yeates, \& Badman}]{STANSBY2020}
Stansby, D., Yeates, A., \& Badman, S.~T. 2020, Journal of Open Source Software, 5, 2732, \dodoi{10.21105/joss.02732}

\bibitem[{Vršnak {et~al.}(2012)Vršnak, Žic, Vrbanec, Temmer, Rollett, Möstl, Veronig, Čalogović, Dumbović, Lulić, Moon, \& Shanmugaraju}]{Vr_nak_2012}
Vršnak, B., Žic, T., Vrbanec, D., {et~al.} 2012, Solar Physics, 285, 295–315, \dodoi{10.1007/s11207-012-0035-4}

\bibitem[{Vršnak {et~al.}(2014)Vršnak, Temmer, Žic, Taktakishvili, Dumbović, Möstl, Veronig, Mays, \& Odstrčil}]{Vr_nak_2014}
Vršnak, B., Temmer, M., Žic, T., {et~al.} 2014, The Astrophysical Journal Supplement Series, 213, 21, \dodoi{10.1088/0067-0049/213/2/21}

\bibitem[{Yamazaki {et~al.}(2024)Yamazaki, Matzka, da~Silva, Kervalishvili, Korte, \& Rauberg}]{Yamazaki2024}
Yamazaki, Y., Matzka, J., da~Silva, M.~V., {et~al.} 2024, \dodoi{10.22541/essoar.171838396.68563140/v1}

\end{thebibliography}
\bibliographystyle{aasjournal}

\newpage
\begin{figure*}
	\includegraphics[width=\textwidth]{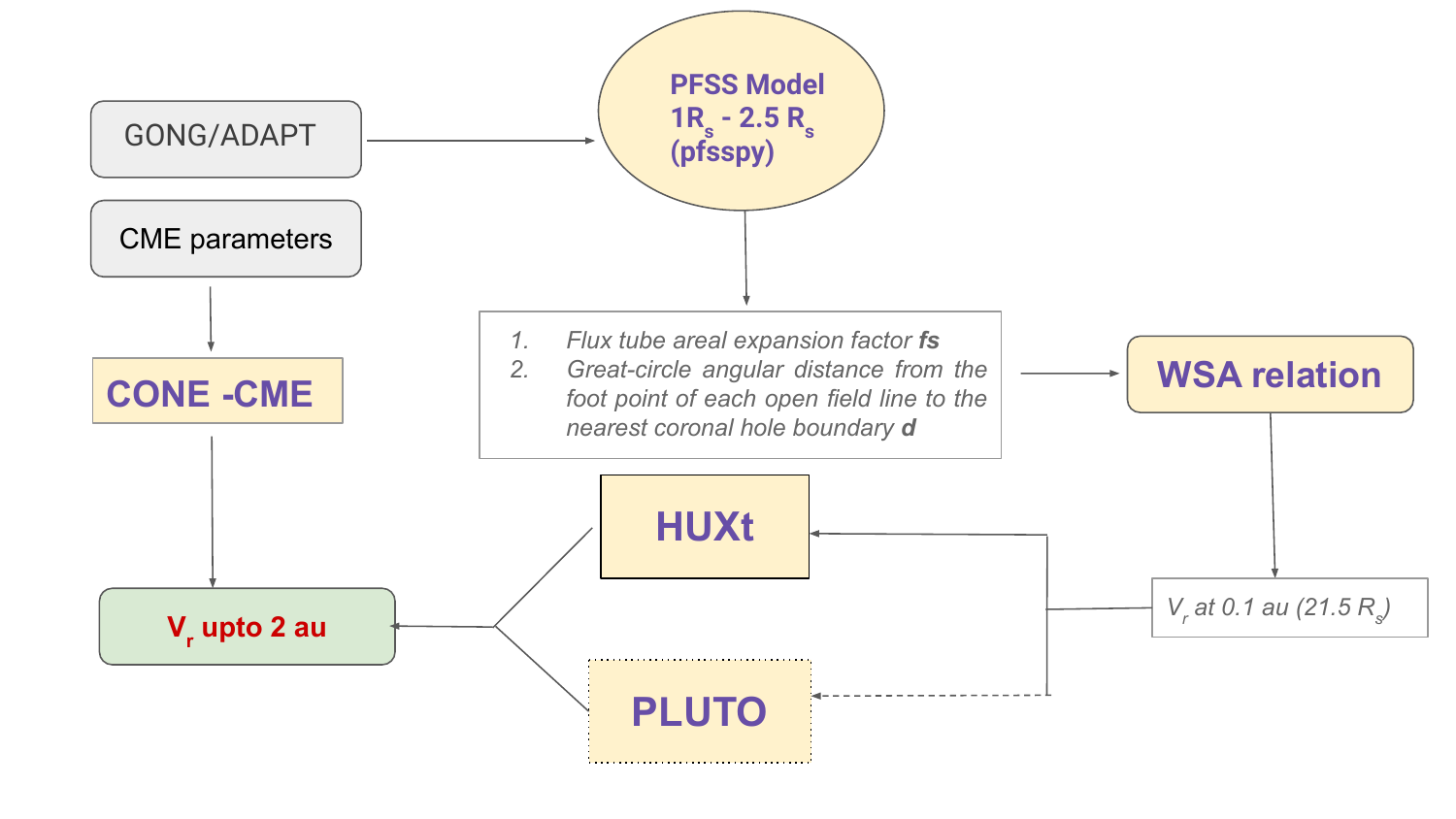}
    \caption{The flow chart depicting the details of the STAR (Solar Transient ARrival) and SWASTi- MHD models,The STAR is based on PFSS coronal model, WSA semi-empirical relation, and HUXt with CONE-CME, whereas the SWASTi model is based on PFSS coronal model, WSA semi-empirical relation, and PLUTO-MHD code with CONE-CME.}
    \label{fig:flowchart}
\end{figure*}

\begin{figure*}
	\includegraphics[width=\textwidth]{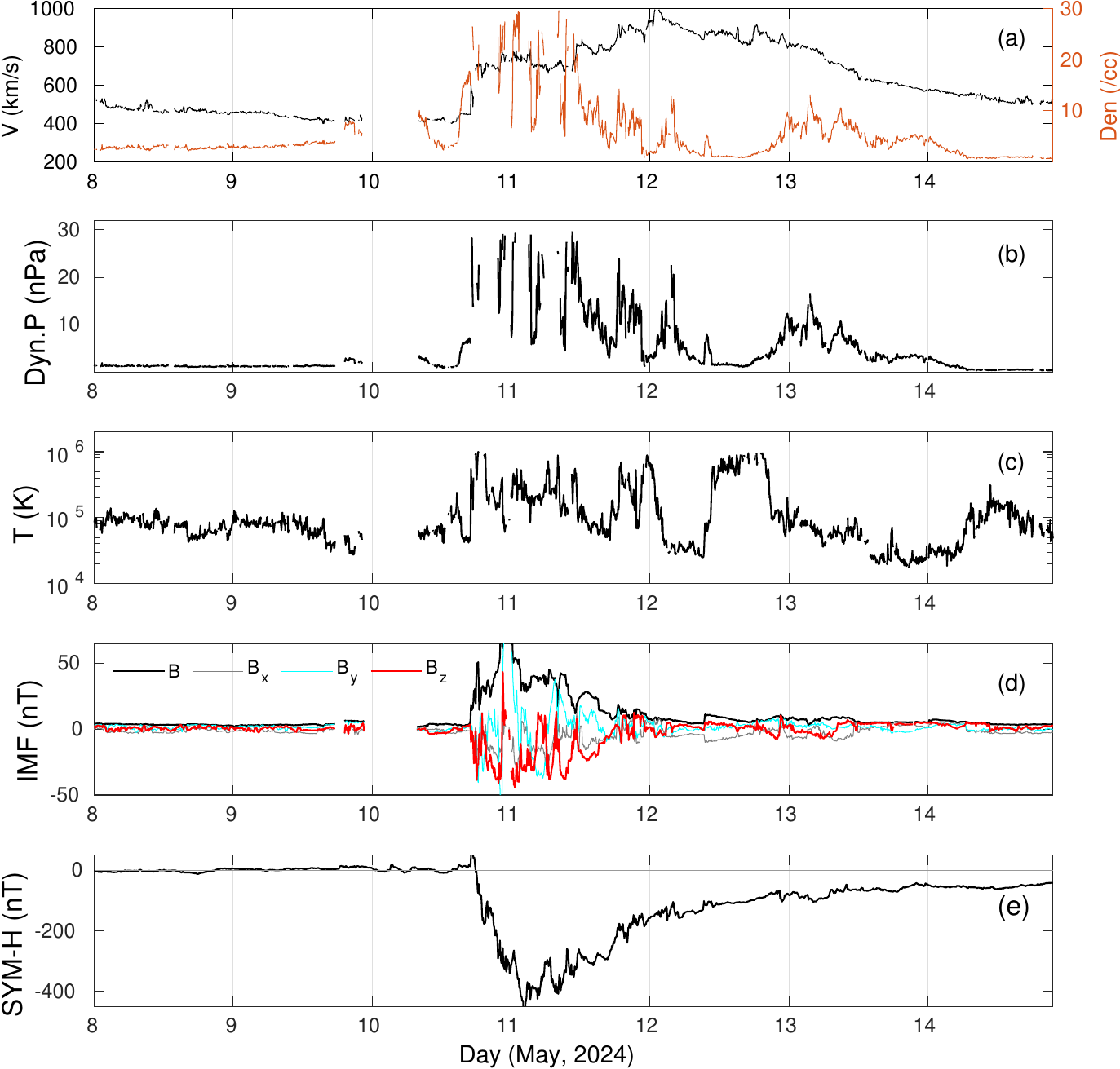}
    \caption{The temporal variations of (a) solar wind speed (left y-axis) and solar wind density (right-y axis), (b) solar wind dynamic pressure (c) Temperature, (d) Interplanetary Magnetic field measured at Sun-Earth L1  and (e) the Sym-H variation at Earth for the period 8-14 May 2024. }
    \label{fig:bkg}
\end{figure*}

\begin{figure*}
	\includegraphics[width=\textwidth]{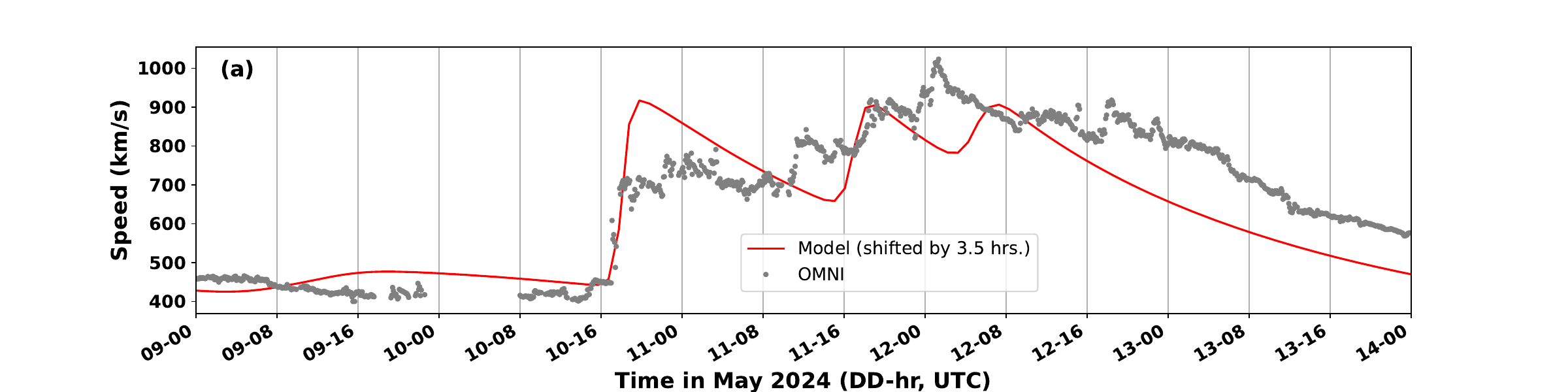}\\
    \includegraphics[width=\textwidth]{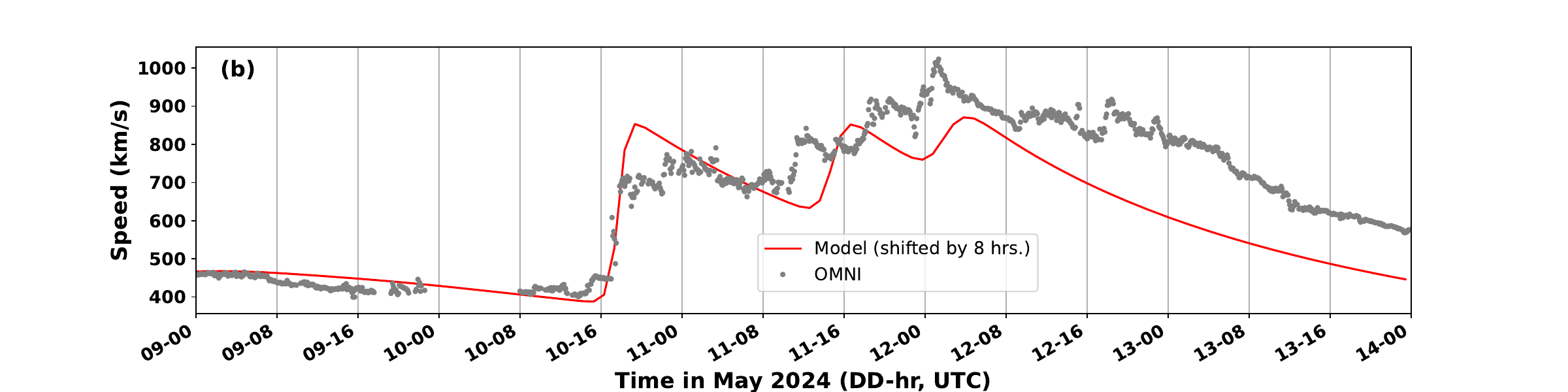}\\
     \includegraphics[width=\textwidth]{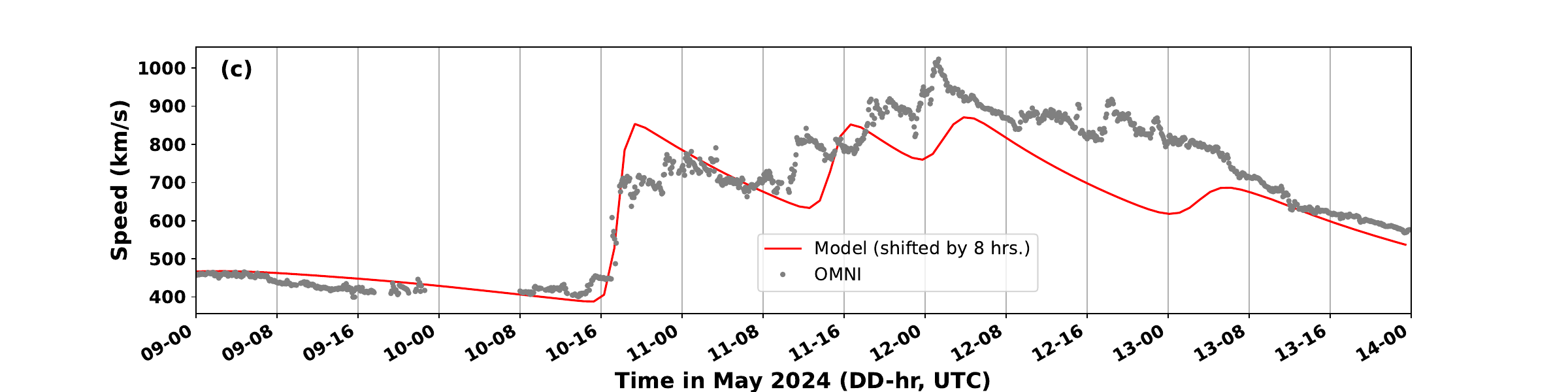}\\
    \caption{The solar wind velocity simulated by STAR module (red lines) compared with actual observations (dots). (a) simulations with GONG  map obtained on 09 May (b) and (c)simulations with GONG  map obtained on 10 May with different CME inputs. }
    \label{fig:gong}
\end{figure*}

\begin{figure*}
	\includegraphics[width=\textwidth]{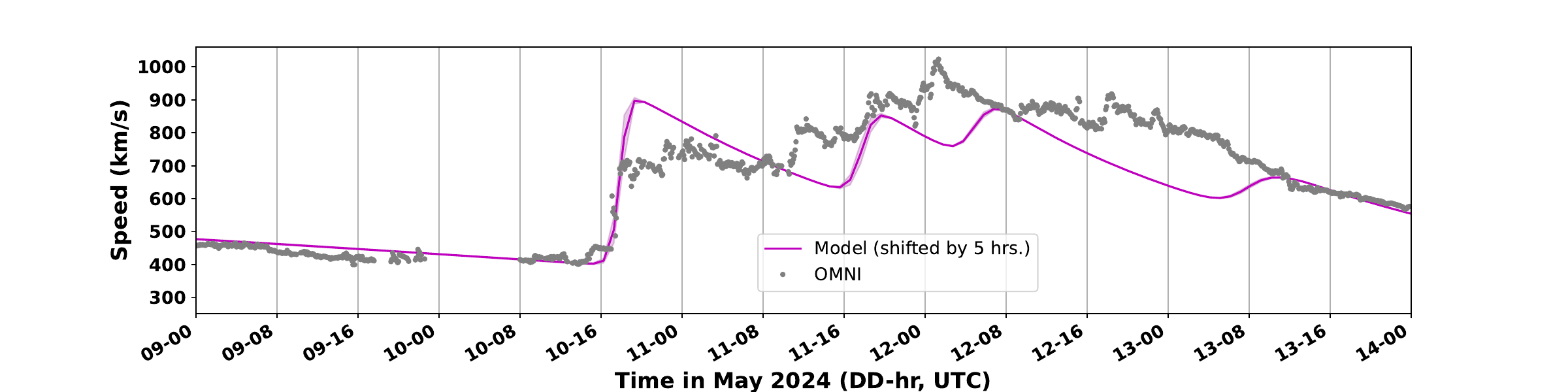}\\
    
   \caption{The solar wind velocity simulated by STAR module using GONG-ADAPT maps compared with actual observations.}
    \label{fig:adapt}
\end{figure*}

\begin{figure*}

	\centering\includegraphics[width=1\textwidth]{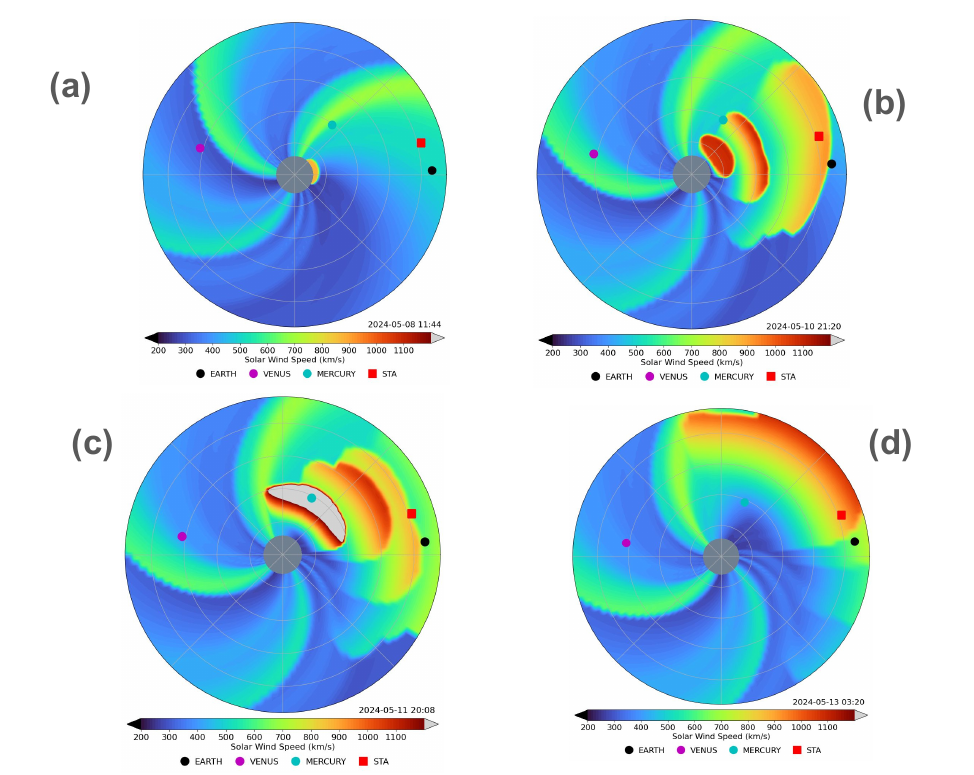}\\
    
    \caption{Snapshots from the STAR simulations. The solar wind velocity is shown in color. Different CMEs can be seen. The positions of Mercury, Venus, Earth, and STEREO-A spacecraft are also shown. The UTC corresponding to each snapshot is given in the figure.}
    \label{fig:huxmaps}
\end{figure*}

\begin{figure*}
    \centering
    \includegraphics[width=0.8\textwidth]{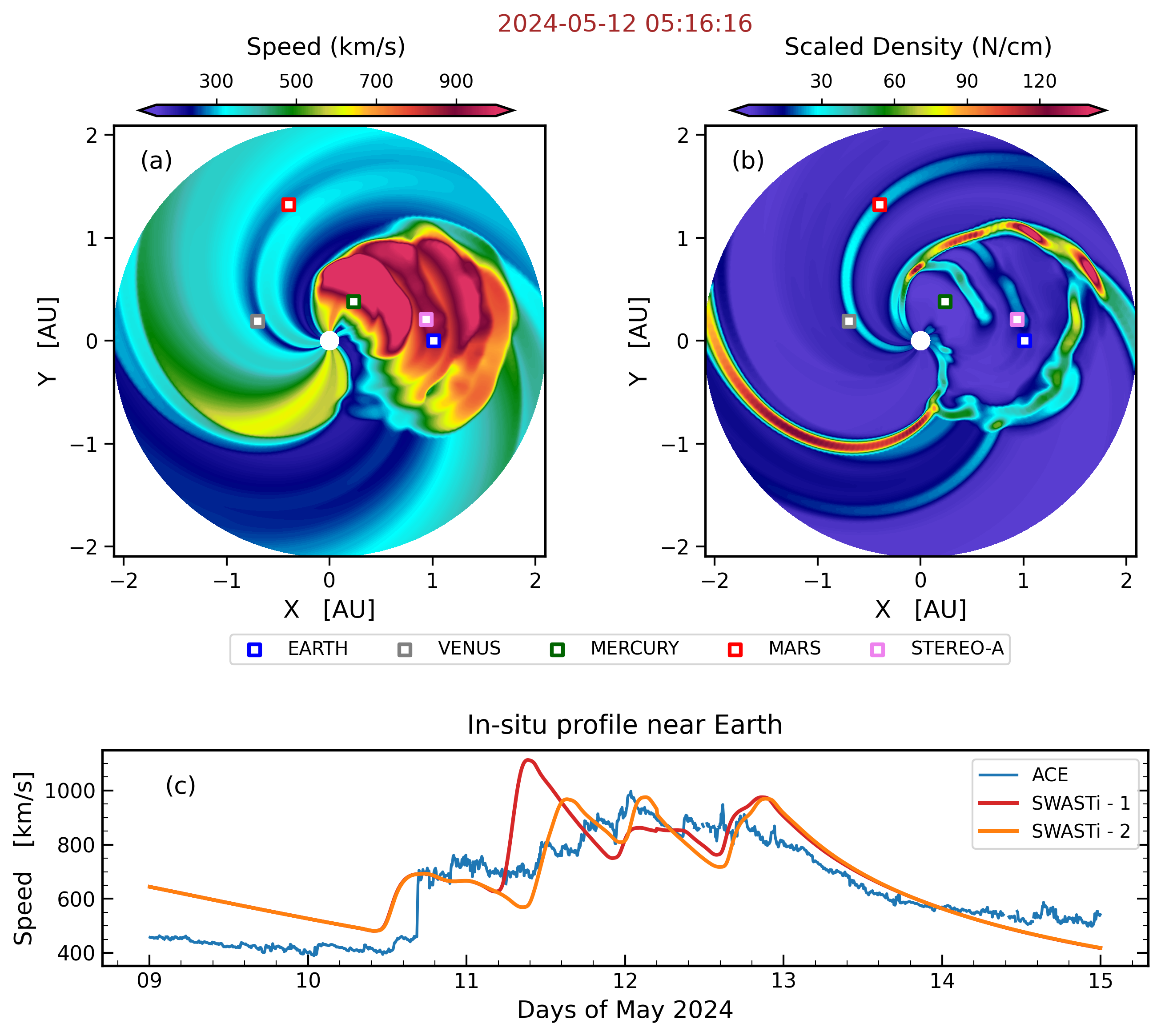}
    \caption{The figure demonstrates the simulation results from the SWASTi framework. Subplots (a) and (b) show a time snapshot, at 2024-05-12T05:16, of the speed and scaled density in the equatorial plane, respectively. Subplot (c) shows the in-situ speed profile near Earth, where the blue line represents the ACE observational data, and the red \& orange lines correspond to the default (1) and tuned (2) SWASTi results, respectively.}
    \label{fig:SWASTi_plot}
\end{figure*}

\begin{figure*}
    \centering
    \includegraphics[width=0.8\textwidth]{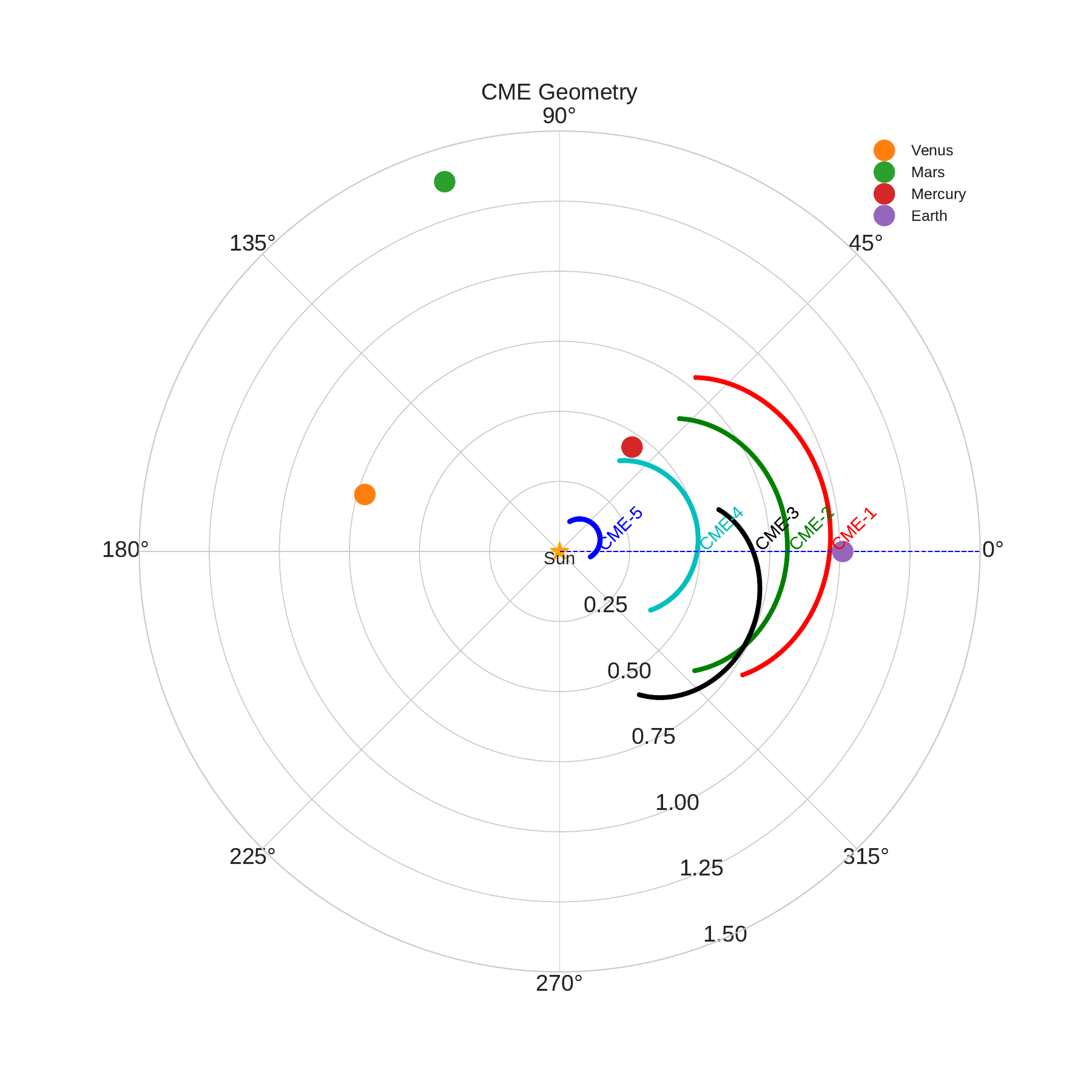}
    \caption{The CME fronts modeled based on DBM model are shown for all five CMEs, the snapshot is at the time of arrival of CME-1 at Earth. Note that CME-CME interaction is not accounted for the CME-propagation.  }
    \label{fig:geometry_dbm}
\end{figure*}

\begin{figure*}
    \centering
    \includegraphics[width=0.9\textwidth]{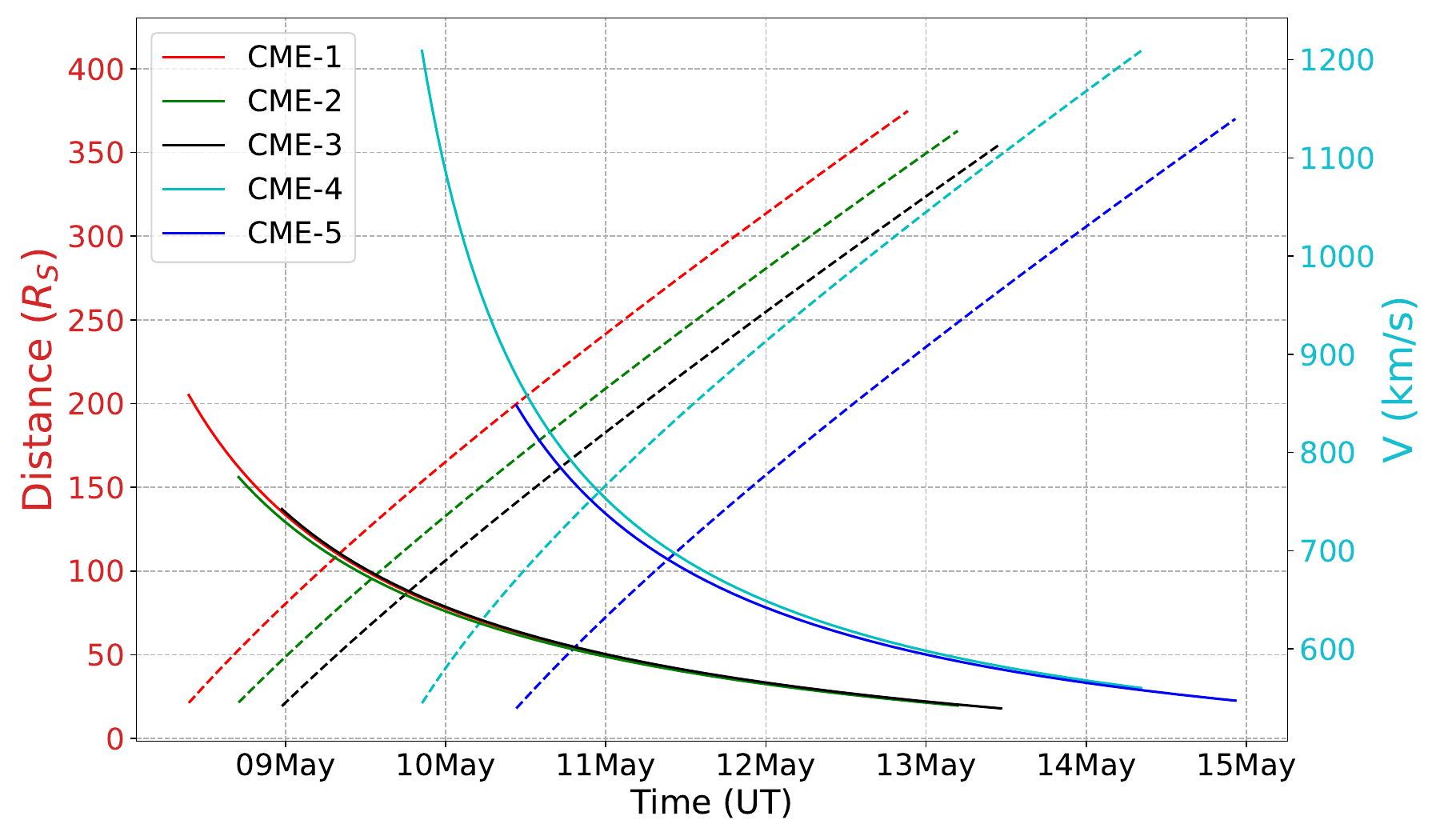}
    \caption{The DBM modeled CME's temporal evolution of velocity (solid lines) and radial distance (dashed lines) of CME from the Sun. The colors indicate the modeled CME and the dashed line indicates radial distance. }
    \label{fig:RT}
\end{figure*}

\begin{figure*}
    \centering
    \includegraphics[width=\textwidth]{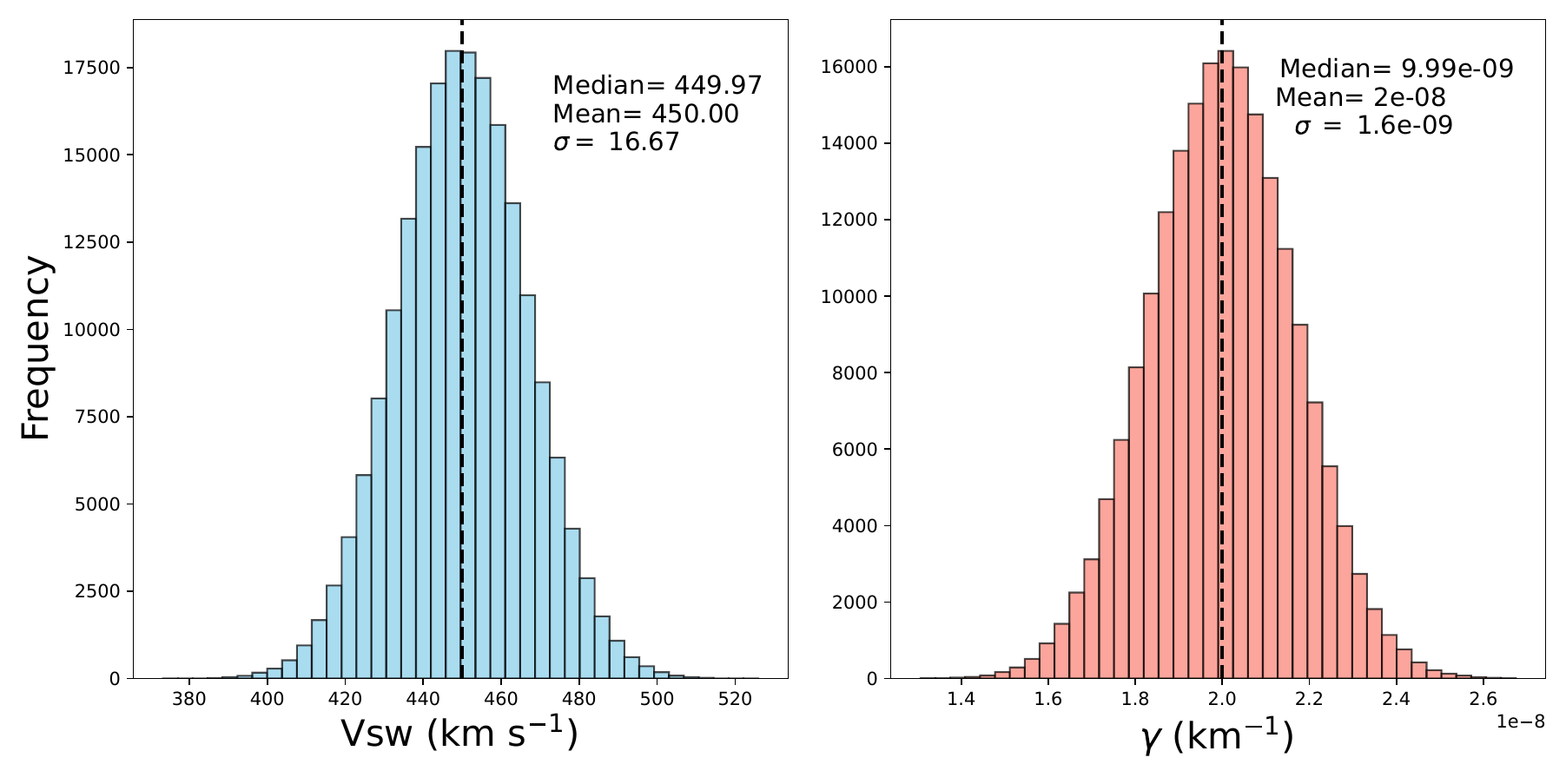}
    \caption{The Gaussian histogram of background solar wind velocity and drag parameter for generated input parameter ensemble for DBM model. For solar wind speed mean is assumed to be 450km/sec and $3\sigma=50$ , whereas for $\gamma$ mean is assumed to be $2 \times 10^{-8}$   and  $3\sigma=0.5 \times 10^{-8}$.}
    \label{fig:hist_input}
\end{figure*}


\begin{figure*}
    \centering
    \includegraphics[width=\textwidth]{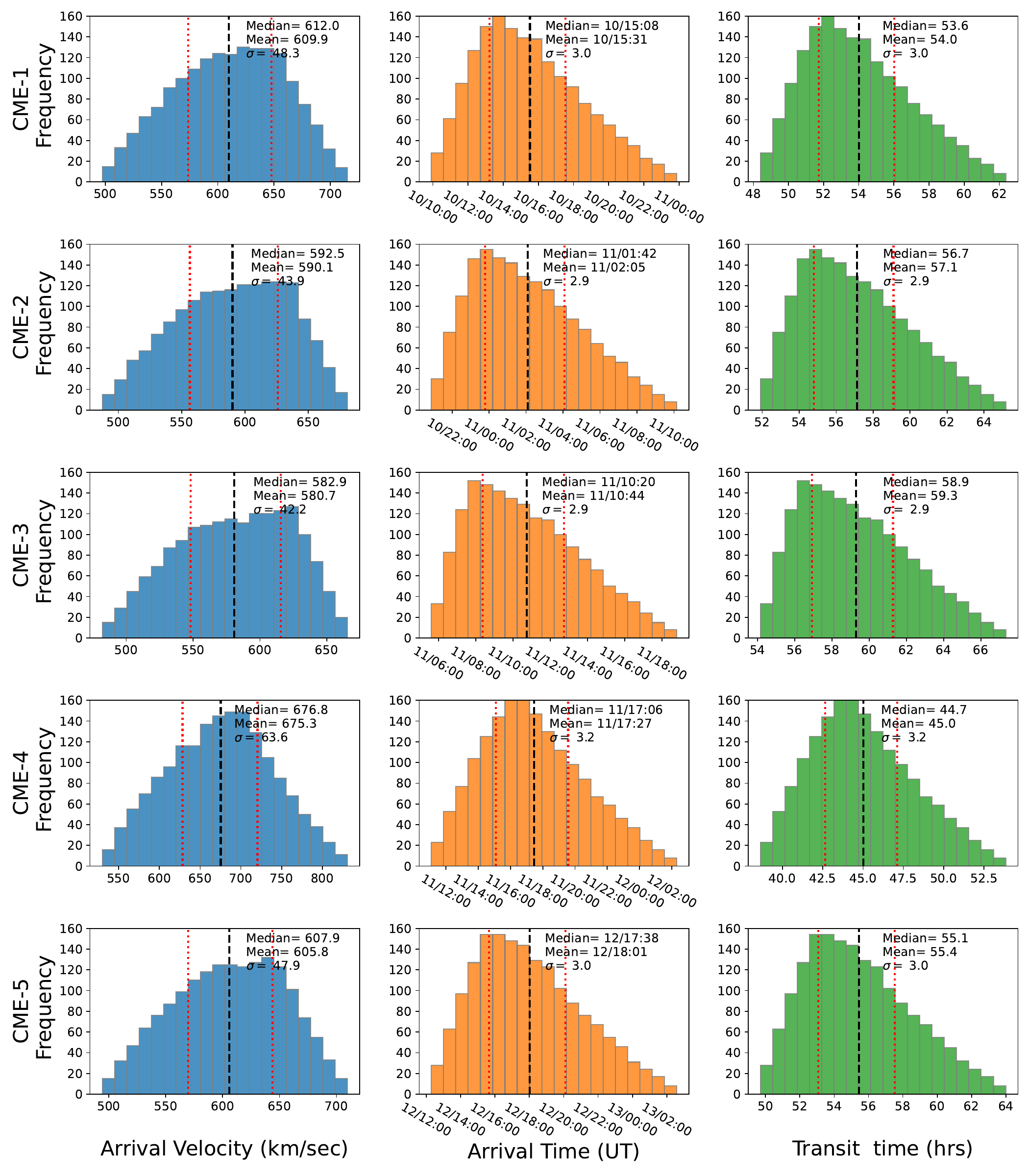}
    \caption{The histogram of the ensemble for each CME is shown for arrival velocity (blue), arrival time (orange), and transit time (green). Each row presents the parameters of the labeled CME.  The frequency indicates the number of events falling in that particular bin of the distribution. The dashed vertical black line marks the mean value of the distribution and the red vertical dotted lines from left to right mark the lower (LQ) and upper(UQ) quartile respectively. CME labels at  the left side indicate the CMEs listed in Table 1.}
    \label{fig:hist_output}
\end{figure*}

\end{document}